\documentclass[aps,                    
prd,                    
showpacs,               
superscriptaddress,    
nofootinbib,            
showkeys,               %
preprintnumbers,        %
floatfix]               
{revtex4}               
\usepackage{graphicx,longtable,amsmath}
\usepackage{natbib}
\usepackage{soul}
\usepackage{xcolor}
\usepackage{xspace}
\usepackage{mathtools}

\numberwithin{equation}{section}
\newcommand{\neff}{\ensuremath{N_{\rm eff}}\xspace}
\newcommand{\tcm}{\ensuremath{T_{\rm cm}}\xspace}
\newcommand{\tdecx}{\ensuremath{T_{\nu_x, {\rm dec}}}\xspace}
\newcommand{\tdece}{\ensuremath{T_{\nu_e, {\rm dec}}}\xspace}
\newcommand{\tdecm}{\ensuremath{T_{\nu_\mu, {\rm dec}}}\xspace}
\newcommand{\tdect}{\ensuremath{T_{\nu_\tau, {\rm dec}}}\xspace}
\newcommand{\tnux}{\ensuremath{T_{\nu_x}}\xspace}
\newcommand{\nue}{\ensuremath{\nu_e}\xspace}
\newcommand{\bnue}{\ensuremath{\overline{\nu}_e}\xspace}
\newcommand{\num}{\ensuremath{\nu_\mu}\xspace}
\newcommand{\nut}{\ensuremath{\nu_\tau}\xspace}
\newcommand{\emu}{\ensuremath{\mu_{e\mu}}\xspace}
\newcommand{\mutau}{\ensuremath{\mu_{\mu\tau}}\xspace}
\newcommand{\etau}{\ensuremath{\mu_{e\tau}}\xspace}

\begin{document}

\title{Majorana Neutrino Magnetic Moment and Neutrino Decoupling in Big Bang Nucleosynthesis}
\author{N. Vassh}
\email{vassh@wisc.edu}
\affiliation{Department of Physics, University of Wisconsin, Madison, WI 53706, USA}
\author{E. Grohs}
\email{egrohs@umich.edu}
\affiliation{Department of Physics, University of Michigan, Ann Arbor, MI 48109, USA}
\author{A.~B. Balantekin}
\email{baha@physics.wisc.edu}
\affiliation{Department of Physics, University of Wisconsin, Madison, WI 53706, USA}
\author{G.~M. Fuller}
\email{gfuller@ucsd.edu}
\affiliation{Department of Physics, University of California, San Diego, La Jolla, CA 92093, USA}

\date{\today}


\begin{abstract}
We examine the physics of the early universe when Majorana neutrinos (\nue, \num, \nut)
possess transition magnetic moments.  These extra couplings beyond the usual
weak interaction couplings alter the way neutrinos decouple from the plasma of
electrons/positrons and photons.  We calculate how transition magnetic moment
couplings modify neutrino decoupling temperatures, and then use a full weak,
strong, and electromagnetic reaction network to compute corresponding changes in
Big Bang Nucleosynthesis abundance yields.  We find that light element abundances and other cosmological parameters are sensitive to magnetic couplings on the order of $10^{-10}\mu_B$. Given the recent analysis of sub--MeV Borexino data which constrains Majorana moments to the order of $10^{-11}\mu_B$ or less, we find that changes in cosmological parameters from magnetic contributions to neutrino decoupling temperatures are below the level of upcoming precision observations.
\end{abstract}

\maketitle

\section{Introduction}

In this paper we explore how the early universe, and the weak decoupling/Big Bang Nucleosynthesis (BBN) epochs in particular, respond to new neutrino sector physics in the form of beyond--the--standard model neutrino magnetic moments.  Small electromagnetic couplings of neutrinos, e.g., magnetic moments, may have effects in the early universe, where neutrinos determine much of the energetics and composition (e.g., isospin). Examinations of the role that neutrino magnetic channels can play in astrophysics and cosmology began long before laboratory experiments, such as GEMMA \cite{Beda}, ever attempted to detect beyond--the--standard model magnetic moments. Any plasma environment in which neutrinos are produced in copious amounts, such as the Sun and supernovae, is susceptible to the possibility of small neutrino magnetic moments having observable consequences. Since neutrinos are a dominant constituent of the early universe during the BBN era, this environment is sensitive to the additional interactions that the magnetic channels provide. For example,  changes in the primordial $^4$He abundance due to $\nu\bar{\nu}$ annihilation into electron--positron pairs were used to provide limitations on the mass and magnetic moment of tau neutrinos before experiment was able to exclude tau neutrino mass of order MeV \cite{KawanoFuller,Grasso}. Similarly, massive sterile neutrinos have the ability to magnetically decay into a light neutrino while injecting a non-thermal photon into the primordial plasma. Such processes alter the primordial abundance yields which can be used to constrain the allowed sterile neutrino mass and magnetic moment parameter space \cite{Kusakabe}. \newline

\noindent Since the magnetic moment interaction always changes the helicity of
the incoming neutrino, Dirac neutrinos in particular have been of interest in
BBN research due to their ability to populate the ``wrong" helicity states. For
instance studies of the spin--flip rate in a primordial magnetic field
\cite{Enqvist} and primordial plasmon decay into $\nu\bar{\nu}$ pairs
\cite{GrassoKolb} are centered around the possibility of generating
right--handed neutrino states. If the neutrino is a Dirac particle, magnetic
interactions between the neutrinos and charged leptons could keep the
right--handed states in thermal, Fermi--Dirac equilibrium with the left--handed
states. The higher energy density would increase \neff (the so--called
effective number of relativistic degrees of freedom measured by cosmic
microwave background (CMB) experiments) from $\sim3$ to $\sim6$ which is in
disagreement with current observations \cite{PlanckXVI}. This has lead many
authors to examine limits on the magnetic moments of Dirac neutrinos by
constraining the production of right--handed states during times earlier than
BBN such as the QCD epoch \cite{Morgan,Fukugita,ElmforsEnqvist,Ayala}. Morgan
\cite{Morgan} and Fukugita et al \cite{Fukugita} use an approximate
neutrino--electron scattering cross section to quantify the production of these
helicity states. Elmfors et al \cite{ElmforsEnqvist} and Ayala et al
\cite{Ayala} perform numerical treatments of the right--handed production rate
by considering the proper photon propagator in an electron--positron plasma.
These analyses then require that these helicity states have decoupled prior to
BBN.  The connection with BBN comes through either comparisons of the expansion
rate and right--handed interaction rate or in the case of \cite{Fukugita}
through constraining the number density of right--handed states. Therefore the
limits on the magnetic moment of neutrinos that these works produce are
necessarily functions of the right--handed decoupling temperature. Fukugita et
al \cite{Fukugita} obtain $\mu_\nu < 7\times 10^{-11}\, \mu_B$ for a
right-handed neutrino decoupling temperature of $T_{\rm dec} \simeq 100$ MeV.
Elmfors et al \cite{ElmforsEnqvist} and Ayala et al \cite{Ayala} obtain
$\mu_\nu < 6.2\times 10^{-11}\, \mu_B$ and $\mu_\nu < 2.9\times 10^{-10}\,
\mu_B$ respectively (also for $T_{\rm dec} \simeq 100$ MeV). However,
limitations on the magnetic interaction of neutrinos which appeal to the need
to avoid right--handed states are only applicable to Dirac neutrinos.  A
right--handed Majorana neutrino behaves as an antineutrino and so would not
cause a sizable increase in the effective number of neutrino species. Since for
this work we will consider the case of Majorana neutrinos, inclusion of the
magnetic channels can keep the neutrinos interacting with the primordial plasma
into the BBN era. This allows us to examine explicitly the connection between
neutrino magnetic moment and BBN observables such as the primordial abundances
and $N_{\text{eff}}$. \newline

\noindent In addition to helicity considerations, magnetic interactions of Majorana neutrinos differ from the Dirac case in that Majorana dipole moments are necessarily transition moments. Dirac neutrinos can have both diagonal and non--diagonal moments, but Majorana neutrino diagonal moments are identically zero \cite{Giunti}.  This implies that for Majorana neutrinos the magnetic channels must occur with flavor changing currents such as $\nu_e + e^- \rightarrow \bar{\nu}_\mu + e^-$ and $\nu_e + \nu_\mu \rightarrow e^+ + e^-$. Constraints on the effective magnetic moment of neutrinos measured by experiments such as TEXONO \cite{Deniz} and GEMMA are not readily applicable to the transition magnetic moments of Majorana neutrinos. Experimental upper bounds need to be converted into limits on transition moments, $\mu_{xy}$, where subscripts $x$ and $y$ denote the neutrino states coupling to the photon. Such constraints were first found in \cite{GrimusMaltoni} using available solar + reactor scattering data from experiments such as SNO, SuperK, ROVNO and MUNU. A recent analysis of sub--MeV Borexino scattering data \cite{Borex} found constraints on the three Majorana transition moments in the mass basis to be $\mu_{ij}\leq [3.1-5.6]\times 10^{-11}\mu_{B}$ \cite{Canas}. These new limits given by Borexino data are stronger than those found from MUNU and TEXONO data, and secondary only to constraints from GEMMA data which for Majorana moments are $\mu_{ij}\leq [2.9-5.0]\times10^{-11}\mu_B$ \cite{Canas}. \newline

\noindent The most restrictive constraint on Dirac and Majorana dipole moments comes from astrophysics by considering how energy loss due to plasmon decay in red giant stars affects the core mass at the helium flash \cite{RaffeltRG}. This analysis was updated using the red--giant branch in the globular cluster M5 and found $\mu_{\nu}\leq 4.5\times 10^{-12}\mu_B$ at the $95\%$ CL \cite{Viaux}. The previous constraint is applicable to the sum of neutrino magnetic and electric dipole moments in the mass basis. When considering Majorana transition moments exclusively, previous astrophysical examinations have mainly focused on the spin flavor precession of solar neutrinos  \cite{BalantekinHatchell,Semikoz,Pulido} and neutrinos in supernovae \cite{AkhmedovLanza,1997PhRvD..55.3265N}. The behavior of active neutrino flavor and spin in dense media is governed by quantum kinetic equations which are difficult to solve in the general case, but become simpler in the homogeneous and isotropic conditions of the early universe \cite{1991NuPhB.349..743B,AkhmedovBerezhiani,1993APh.....1..165R,2005PhRvD..71i3004S,2007JPhG...34...47B,2013PhRvD..87k3010V,2013PrPNP..71..162B,Gouvea,2014PhRvD..90l5040S,VFC:QKE,2015PhLB..747...27C}. In contrast to the solar and supernovae environments, the literature on the role of Majorana neutrino transition moments in BBN is sparse. To the best of our knowledge the only investigation of Majorana moments in the early universe appeals to an active--sterile transition moment \cite{Pastor} in the presence of a primordial magnetic field.  \newline

\noindent  In this work, we examine effects of the Majorana neutrino transition moments in the early universe without invoking sterile neutrinos or primordial magnetic fields. If its transition magnetic moments are large enough, a given active neutrino species can remain coupled to the electromagnetic plasma into the BBN epoch. In the early universe, a Majorana neutrino can magnetically interact with electrons and positrons as well as photons and other neutrinos. However magnetic neutrino--neutrino scattering and interactions between neutrinos and photons (such as neutrino Compton scattering) are proportional to the fourth power of the magnetic moment \cite{Aydin} and so are suppressed relative to interactions with electrons and positrons which are proportional to the second power of the magnetic moment. Here we explore effects on the primordial abundances when enhanced interactions between electrons/positrons and Majorana neutrinos via magnetic scattering and annihilation channels are taken into account. In section II we demonstrate how magnetic neutrino--electron scattering can play a significant role in a relatively low temperature environment such as the early universe. In section III we find the neutrino interaction rate for the case of thermal equilibrium by introducing expressions for the thermally averaged cross section times Moller velocity which make use of Fermi--Dirac statistics. We use these interaction rates to find neutrino decoupling temperatures as a function of the transition magnetic moments $\mu_{e\mu}$, $\mu_{e\tau}$, and $\mu_{\mu\tau}$. In section IV we explore the corresponding change in the predicted abundances and $N_{\text{eff}}$ when these decoupling temperatures are implemented in a modified version of the Wagoner--Kawano (WK) code \cite{Wagoner69, SKM93}. We conclude in section V.  Appendix A discusses the inverse Debye screening length at all temperatures for an electron--positron plasma which is used here to demonstrate that the magnetic scattering cross section is finite.  Appendix B shows the derivation of the interaction rate expressions employed here which are functions of the cross section. All relevant cross sections are listed in Appendix C.  Throughout this work, we use natural units where $\hbar=c=k_B=1$.

\section{Low Temperature Enhancement to Magnetic Scattering}

It is well known that for neutrino--electron scattering the magnetic channel can dominate over the weak channel when the recoil energy of the outgoing electron is sufficiently small \cite{Balantekin}. Experiments which aim to measure the magnetic moment of electron antineutrinos from reactors, such as TEXONO and GEMMA, exploit this property to obtain bounds on neutrino moments. By continuously pushing their threshold energies lower, these experiments are able to correspondingly obtain more restrictive upper bounds on the measured effective magnetic moment given by

\begin{equation}
\label{101}
\mu_{\bar{\nu}_e,\text{eff}}^2 = \sum_j \bigg|\sum_i U_{ei}e^{-iE_i L}\mu_{ij}\bigg|^2,
\end{equation}

\noindent where $L$ is the distance between the neutrino source and detector, $U_{ei}$ is the appropriate vacuum PMNS matrix element, and $\mu_{ij}$ are the values of neutrino magnetic moments in the mass eigenstate basis since this is the appropriate interaction basis for the magnetic channel. The sum over $j$ is performed outside the square in Eq. \eqref{101} since reactor experiments which search for neutrino magnetic moment do not measure the final state of the scattered neutrino. The most recent upper limit given by TEXONO is $\mu_{\bar{\nu}_e,\text{eff}} < 2.2\times 10^{-10}\mu_B$ at the $90\%$ confidence level \cite{Deniz}. The lowest experimental upper limit comes from GEMMA with $\mu_{\bar{\nu}_e,\text{eff}} < 2.9 \times 10^{-11}\mu_B$ at the 90$\%$ confidence level using a threshold energy of $\sim$2.8 keV \cite{Beda}. As previously mentioned, these constraints must be translated into bounds on the transition magnetic moments. The low energy enhancement of the magnetic scattering channel over the weak channel can play a role in environments other than the laboratory such as the early universe during big bang nucleosynthesis. To demonstrate this we examine the low temperature behavior of the magnetic neutrino--electron scattering cross section:

\begin{equation}
\label{1}
\sigma(s) =  \frac{\pi \alpha^2 \mu^2_{\nu}}{m_e^2}\left[\frac{|t_{\text{max}}|}{s-m_e^2}-\frac{s-m_e^2 }{s}+\ln{\frac{\left(s-m_e^2\right)^2}{s\, |t_{\text{max}}|}}\right],
\end{equation} 

\noindent where $\alpha$ is the fine structure constant and $\mu_{\nu}$ is the neutrino's effective magnetic moment in units of Bohr magnetons. Here $s$ is the Mandelstam variable (related to the incoming neutrino energy in the electron's rest frame by $s=m_e^2 + 2m_eE_{\nu}$) and $|t_{\text{max}}|$ is the magnitude of the upper bound of the Mandelstam variable $t$ (related to the minimum value of the electron recoil energy, $T_{e,\text{min}}$, in the electron's rest frame by $t_{\text{max}}=-2m_eT_{e,\text{min}}$). In this work we cut off the infrared divergence by using $t_{\text{max}} = -2m_e\left(\sqrt{k_{sc}^2 + m_e^2}-m_e\right)$ where $k_{sc}$ is the inverse Debye screening length. For a homogeneous and isotropic electron--positron plasma in the absence of an external magnetic field, dynamic screening need not be considered and the static inverse screening length can be found to be  

\begin{equation}
\label{2}
k_{sc}^2 = \frac{4\alpha}{\pi T}\int_0^{\infty} dp\,p^2 \frac{1}{1+\cosh(E/T)},
\end{equation}

\noindent where we have taken the chemical potential of electrons and positrons to be negligible (see Appendix A). In Fig. 1 we compare the weak and magnetic cross sections for electron neutrino scattering with electrons and positrons. Fig. 1 shows explicitly that at low temperature the magnetic channel can dominate over the weak channel particularly for the low energy portion of the thermal distribution of the neutrino. \newline

\begin{figure}[!ht]
\begin{center}
\includegraphics[scale=0.725]{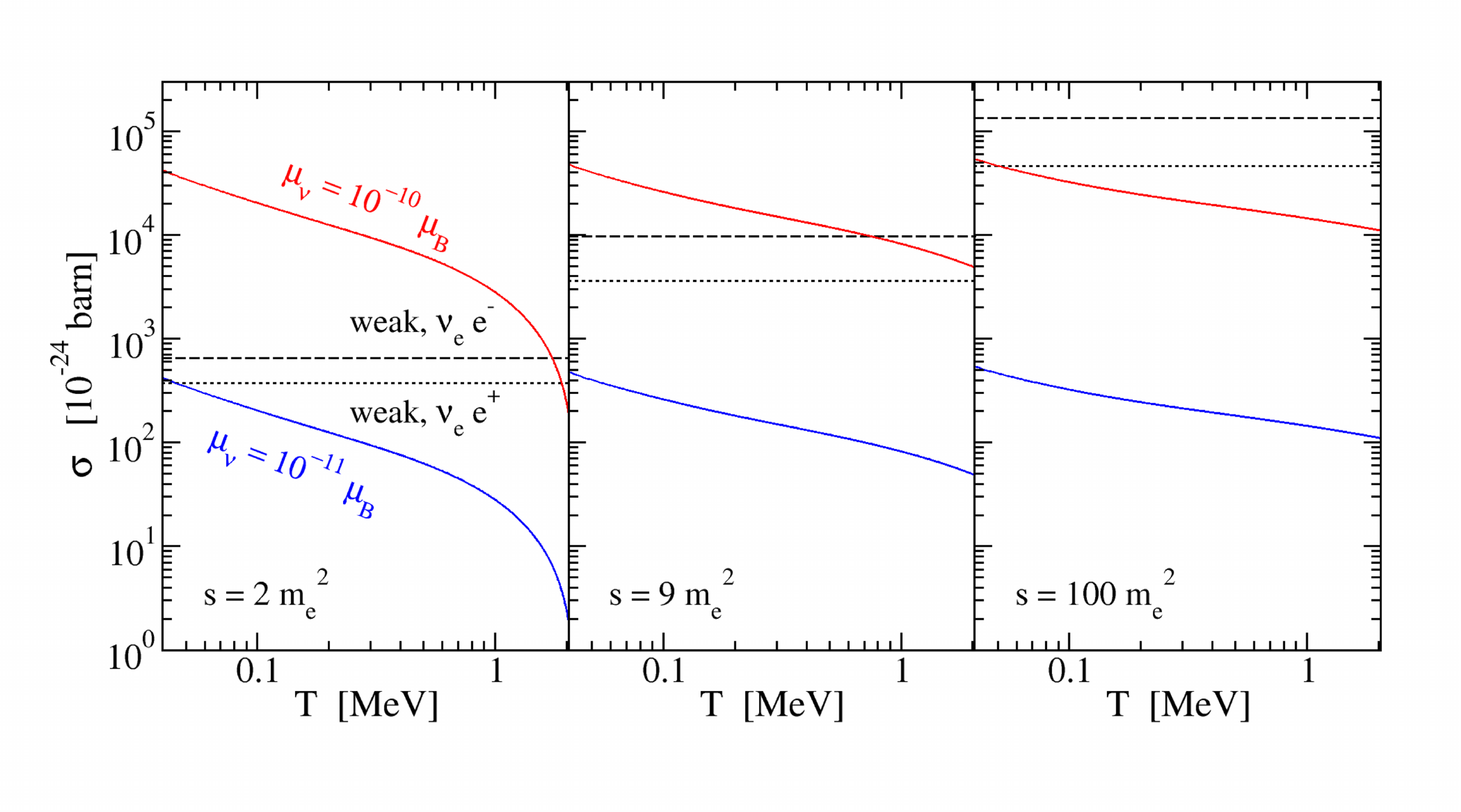}
\caption{(Color online) The weak and magnetic scattering cross sections as a function of temperature for different values of the invariant variable $s$. The weak cross sections for $\nu_e - e^-$ and $\nu_e - e^+$ scattering are represented by the black dashed and black dotted lines respectively. The magnetic cross sections for neutrino--electron (or positron) scattering are shown as the solid lines for two possible effective magnetic moments (red- $10^{-10}\mu_B$ and blue - $10^{-11}\mu_B$).}
\label{fig:fig1}
\end{center}
\end{figure}

\noindent The treatment of the divergent behavior of the scattering cross section $\sigma \propto \ln\left(q_{\text{max}}^2/q_{\text{min}}^2\right)$ is what led to the reexamination of the result of Morgan \cite{Morgan} in Refs.\ \cite{Fukugita,ElmforsEnqvist,Ayala}. Fukugita et al \cite{Fukugita} take the minimum photon momenta transfer to be $q_{\text{min}}\rightarrow 2\pi l_{D}^{-1}$ where $l_{D}=(T/4\pi n\alpha)^{1/2}$ is the Debye screening length in the classical limit. Elmfors et al \cite{ElmforsEnqvist} and Ayala et al \cite{Ayala} perform more proper numerical treatments of the plasma effects on the interaction rate by modifying the photon propagator explicitly. Simple Debye screening of the Coulomb divergence using the inverse screening length will only slightly underestimate the contribution of the photon's longitudinal mode to the cross section \cite{BraatenYuan}. Here we extend the approach of \cite{Fukugita} by implementing the proper expression for the inverse Debye length, Eq. \eqref{2}, at all temperatures. \newline

\section{Decoupling Temperature and Interaction Rate Calculation}

To examine magnetic effects of massless Majorana neutrinos in BBN, flavor changing currents must be considered. To treat this behavior correctly would require solving the Boltzmann and quantum kinetic equations \cite{1991NuPhB.349..743B,AkhmedovBerezhiani,1993APh.....1..165R,2005PhRvD..71i3004S,2007JPhG...34...47B,2013PhRvD..87k3010V,2013PrPNP..71..162B,Gouvea,2014PhRvD..90l5040S,VFC:QKE,2015PhLB..747...27C} to properly evolve the number densities of each neutrino flavor, while simultaneously following a BBN network. A code in development which incorporates neutrino transport when calculating primordial abundances and other cosmological parameters, \texttt{BURST} \cite{numr, 5pts}, will eventually be able to properly treat flavor changing currents while self-consistently calculating the primordial abundances. For this work, we proceed by approximating neutrino decoupling as a sharp event for each flavor. \newline

\noindent Within the decoupling temperature approximation we treat flavor changing processes such as scattering, e.g., $\nu_e + e^- \rightarrow \bar{\nu}_\mu + e^-$, and annihilation, e.g., $\nu_e + \num \rightarrow e^+ + e^-$, as equilibrium channels and take $n_{\nu_e}\approx n_{\nu_\mu}\approx n_{\nu_\tau}$. This means the flavor dependence for the magnetic rates only appears in the magnetic moment contained in the cross sections. With the equilibrium approximation of equivalent number densities, the magnetic flavor changing rates can be added directly to the weak flavor preserving rates in order to obtain the total interaction rate $\Gamma = \Gamma_{\text{scatt}}^{\text{weak}} + \Gamma_{\text{scatt}}^{\text{mag}}+\Gamma_{\text{ann}}^{\text{weak}}+\Gamma_{\text{ann}}^{\text{mag}}$ for an incoming neutrino. To find the decoupling temperature, the total interaction rate must be compared with the Hubble expansion rate:

\begin{equation}
H=\sqrt{\frac{8\pi}{3m_{\rm pl}^2}\rho},
\end{equation}

\noindent where $m_{\rm pl}$ is the Planck mass, and $\rho$ is the energy density. A particle species can be roughly considered to be coupled when $\Gamma \geq  H$ or decoupled when $\Gamma < H$.  The decoupling temperature is then the temperature at which the approximate transition from coupled to decoupled occurs \cite{K&T}, and denoted \tdecx for species $x$.  Here we consider the interactions of neutrinos with electrons and positrons. This gives $\Gamma_{\text{scatt}} = n_{e}\left<\sigma v_{\text{Mol}}\right>_{\text{scatt}}$ and $\Gamma_{\text{ann}} = n_{\nu}\left<\sigma v_{\text{Mol}}\right>_{\text{ann}}$ when the target particle is an electron/positron or an antineutrino respectively. The thermally averaged cross section times Moller velocity for the case of thermal equilibrium is given by

\begin{equation}
\label{4}
\left<\sigma v_{\text{Mol}}\right> = \frac{\int \sigma v_{\text{Mol}} \frac{d^3 p_1}{1+e^{E_1 /T}} \frac{d^3 p_2}{1+e^{E_2 /T}}} {\int \frac{d^3 p_1}{1+e^{E_1 /T}} \int \frac{d^3 p_2}{1+e^{E_2 /T}}},
\end{equation}

\noindent where $E_1$ and $E_2$ are the energies of the two incoming particles.  In order to evaluate Eq. \eqref{4} in the comoving frame, we adopt the approach of Gondolo and Gelmini \cite{Gondolo}, but introduce expressions which make use of the proper Fermi--Dirac distributions (see Appendix B). For massless neutrinos interacting via the annihilation channel the thermally averaged cross section can be written as

\begin{equation}
\label{9}
\left<\sigma v_{\text{Mol}}\right>_{\text{ann}} = \frac{4\pi^2 T^2}{n_\nu^2}\int_{4m_e^2} ^\infty \sigma s ds \int_{\sqrt{s}/T} ^\infty dx \frac{e^{-x}}{1-e^{-x}}\left[\frac{\sqrt{x^2 -s/T^2}}{2}+\ln\left(\frac{1+e^{-\frac{x+\sqrt{x^2 - s/T^2}}{2}}}{1+e^{-\frac{x-\sqrt{x^2 - s/T^2}}{2}}}\right)\right].
\end{equation}

\noindent For the scattering channel we obtain

\begin{align}
\label{10}
  \left<\sigma v_{\text{Mol}}\right>_{\text{scatt}} = \frac{4\pi^2 T^2}{n_\nu n_e}
  &\int_{2m_e^2} ^\infty \sigma \left(s-m_e^2 \right) ds
  \int_{\sqrt{s}/T} ^\infty dx \frac{e^{-x}}{1-e^{-x}}\\
  \nonumber\times &\left[\frac{\sqrt{x^2 -s/T^2}\sqrt{1-2m_e^2 /s}}{2}
  +\ln\left(\frac{1+e^{-\frac{x+\sqrt{x^2 - s/T^2}\sqrt{1-2m_e^2 /s}}{2}}}
  {1+e^{-\frac{x-\sqrt{x^2 - s/T^2}\sqrt{1-2m_e^2 /s}}{2}}}\right)\right],
\end{align}

\noindent and so the interaction rates can now be found by numerical evaluation of the previous two dimensional integrals using the cross sections as functions of $s$ given in Appendix C. \newline

\noindent In figure 2, we show the rates for weak and
magnetic scattering on both electrons and positrons. The low temperature enhancement of the magnetic scattering channel over the weak channel discussed in section II is evident. For a magnetic moment of $10^{-10}\mu_{B}$ the magnetic scattering rate dominates over the weak rate starting at a temperature of about $0.4$ MeV. Figure 3 shows the weak and magnetic annihilation rates.  Clearly magnetic moments on the order of $10^{-10} \mu_B$ have less influence on the annihilation channel as compared to the scattering channel. \newline 

\begin{figure}[!ht]
\begin{center}
\includegraphics[scale=0.525]{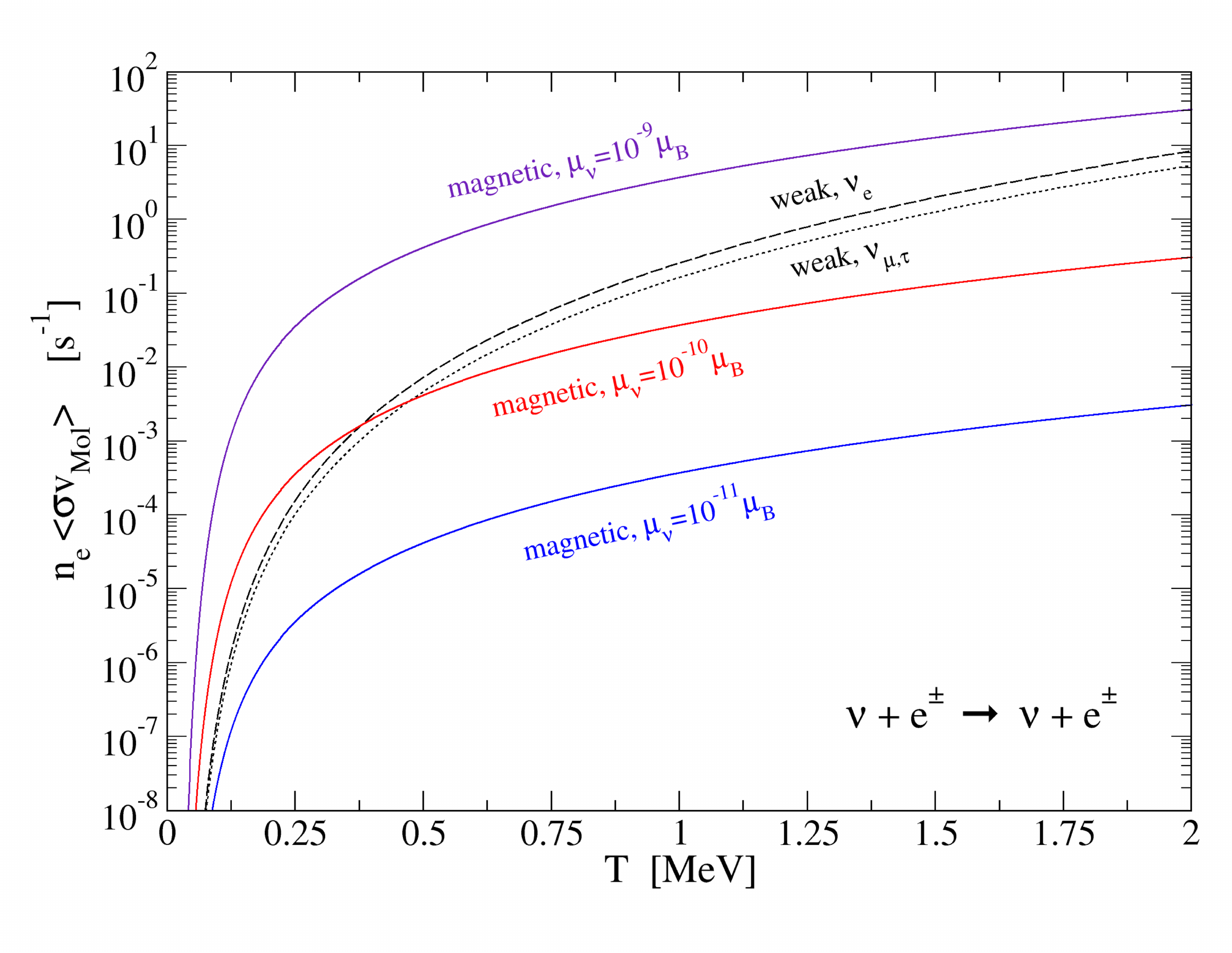}
\caption{(Color online) The weak and magnetic rates for neutrino scattering from electrons and positrons are given as a function of temperature. The weak channel scattering rates are shown as black dashed (for electron neutrinos) and black dotted (for muon and tau neutrinos). Magnetic scattering rates for three possible values of the effective magnetic moment are represented by the solid lines (indigo - $10^{-9}\mu_B$, red - $10^{-10}\mu_B$, and blue - $10^{-11}\mu_B$). }
\label{fig:fig2}
\end{center}
\end{figure}

\begin{figure}[!ht]
\begin{center}
\includegraphics[scale=0.535]{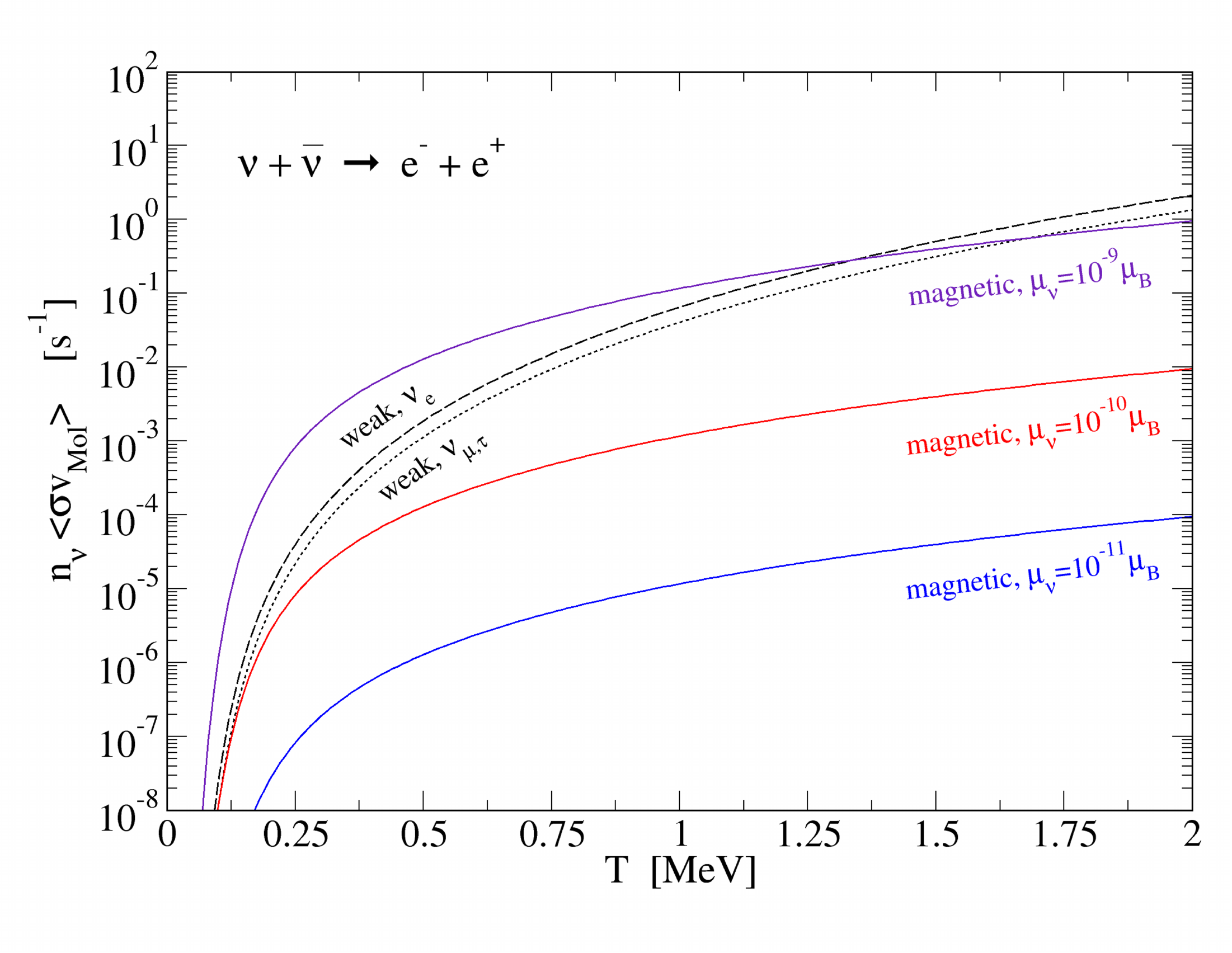}
\caption{(Color online) The weak and magnetic rates for $\nu\bar{\nu}$ annihilation into $e^+ e^-$ pairs. The line assignments are the same as in Fig. 2.}
\label{fig:fig3}
\end{center}
\end{figure}

\noindent To investigate the dependence of the decoupling
temperatures on neutrino transition moments, we introduce
effective magnetic moments \footnote{It should be noted that $\mu_{xy}^2 =
\mu_{xy}\mu_{xy}^* = \mu_{yx}^2$ since the Majorana neutrino magnetic moment
matrix is antisymmetric with $\mu_{xy} = - \mu_{yx} = -\mu_{xy}^*$
\cite{Balantekin} where in the flavor basis these transition moments are given
by $\mu_{xy}^2 = \left|\sum_{i,j}U_{xi}U_{yj}^{\dagger}\mu_{ij}\right|^2$. It is also useful to note that the effective Majorana moments defined in Eq. \eqref{11} have recently been shown to fulfill the triangle inequality $\mu_{\tau,\text{eff}}^2 \leq \mu_{e,\text{eff}}^2+\mu_{\mu,\text{eff}}^2$ \cite{Frere}, although we do not make use of this property here since primordial abundances are found as a function of transition moments. }:

\begin{eqnarray}\label{11}
  \mu_{e,\text{eff}}^2 &=& \mu_{e\mu}^2 + \mu_{e\tau}^2, \\ \nonumber
  \mu_{\mu,\text{eff}}^2 &=& \mu_{\mu e}^2 + \mu_{\mu\tau}^2, \\ \nonumber
  \mu_{\tau,\text{eff}}^2 &=& \mu_{\tau e}^2 + \mu_{\tau\mu}^2. \\ \nonumber
\end{eqnarray}  

\begin{figure}[!ht]
\begin{center}
\includegraphics[scale=0.554]{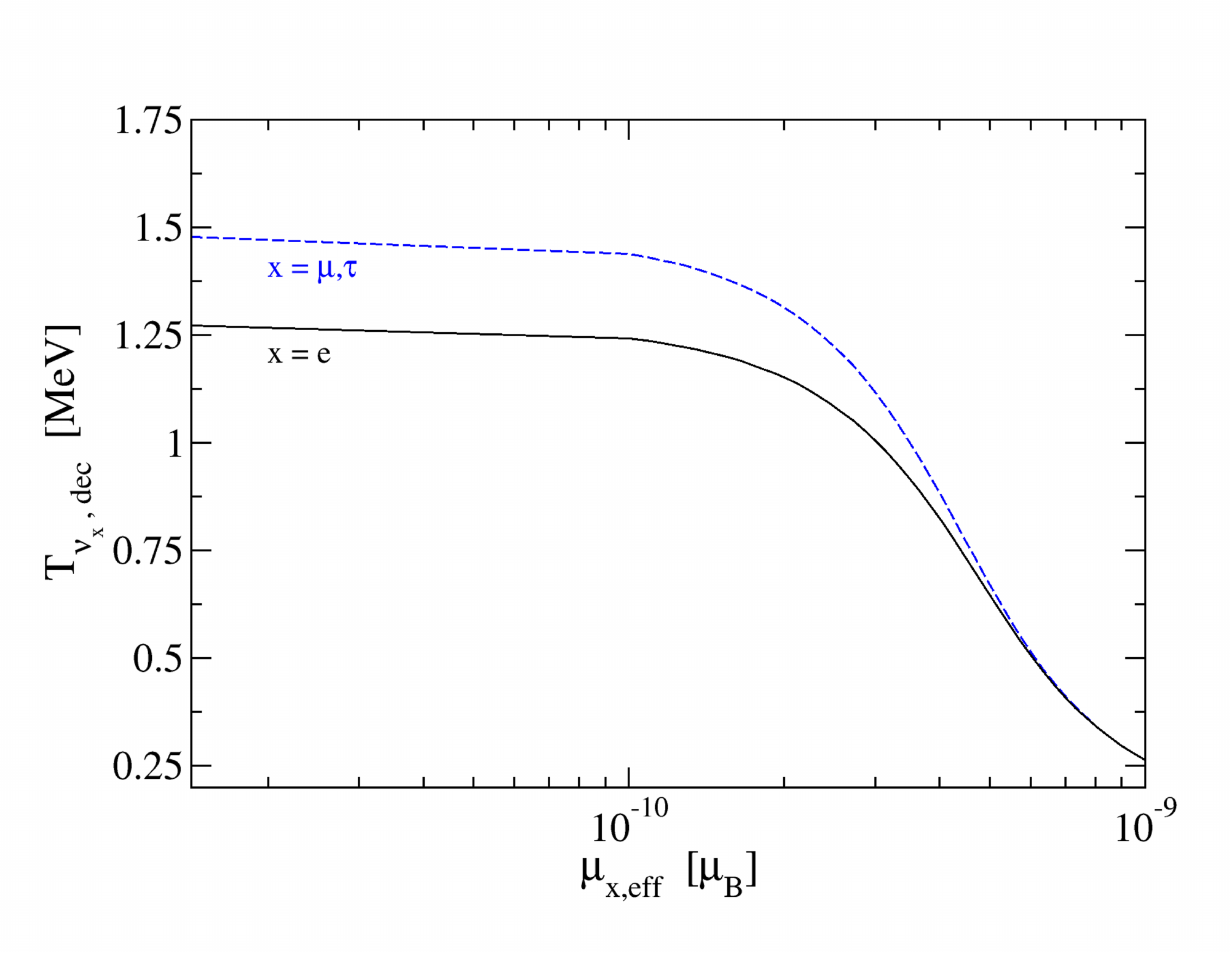}
\caption{(Color online) Decoupling temperature of the three flavors as a
function of their effective magnetic moment. The black solid line
represents electron neutrinos which undergo weak decoupling at $\sim 1.27$ MeV.
The blue dashed line represents muon and tau neutrinos which decouple weakly at
$\sim 1.48$ MeV.} \label{fig:fig4}
\end{center}
\end{figure}

\noindent The effective transition moments defined above are not
identical to the effective electron antineutrino magnetic moment measured by
reactor based experiments such as GEMMA. Here we assume that the incoming and
outgoing massless neutrinos populate definite flavor states in the early
universe. Fig. \ref{fig:fig4} shows the decoupling temperatures for electron
and mu/tau neutrinos as a function of effective magnetic moment.  For the range
$\mu_{x,\text{eff}}\simeq (2-9) \times 10^{-11} \mu_B$, the magnetic
interaction has a small effect on the decoupling temperature.  The magnetic
channels start to play a more significant role when $\mu_{x,\text{eff}} >
10^{-10} \mu_B$ (roughly expected from an examination of the rates in Fig. 2).  If
the \mutau transition magnetic moment is sufficiently high in comparison to the
transition magnetic moments with a \nue component, the \num and \nut species
could remain coupled to the electron--positron plasma longer than the \nue.  
\newline

\section{Changes in Primordial Abundances and $N_{\text{eff}}$}

The WK code assumes at the start of the computation that all flavors of
neutrinos have already decoupled from the plasma.  To alter the
neutrino--decoupling epoch, we modify the summed energy density of neutrinos
and antineutrinos such that

\begin{equation}
\label{12}
\rho_{\nu_x} = \frac{7}{8}\frac{\pi^2}{15}T_{\nu_x}^4,
\end{equation}
\noindent where the temperature of neutrino species $x$, $T_{\nu_x}$, is not
necessarily related to the inverse of the scale factor. We define the comoving
temperature parameter as an energy scale such that the product $\tcm a$ is a
comoving invariant, where $a$ is the scale factor.  To calculate \tcm, we set
\tcm equal to the plasma temperature at an early epoch where the photons,
electrons, positrons, and all flavors of neutrinos and antineutrinos are in
thermal equilibrium.  If we denote this epoch with the subscript $h$, we obtain
the expression $\tcm=T_h(a_h/a)$.  The temperature of each neutrino species
evolves as:

\begin{equation}
\label{14}
  T_{\nu_x} = \begin{dcases}
    T_\gamma&\text{if}\quad\tnux>\tdecx\\
    \displaystyle\frac{{a_{\rm dec}}}{a}\,\tdecx&\text{if}\quad\tnux<\tdecx
  \end{dcases},
\end{equation}
\noindent where $a_{\rm dec}$ is the value of the scale factor at the time of $\nu_x$ decoupling.  Eq. \eqref{14} states that \tnux is equal to the plasma temperature until the decoupling epoch, after which \tnux scales with the comoving temperature.\newline

\noindent The late neutrino decoupling epoch necessitates a modification to the WK temperature derivative.  To calculate the plasma temperature derivative in the presence of coupled neutrinos, we need the quantities:

\begin{align}
\label{13}
p_{\nu_x} &= \frac{\rho_{\nu_x}}{3},\\
\frac{d\rho_{\nu_x}}{dT_\gamma} &= \frac{4\rho_{\nu_x}}{T_\gamma},
\end{align}
\noindent where $p$ is the pressure, and $d\rho/dT_\gamma$ is the temperature derivative of the energy density.  Initially, all three neutrino species stay coupled to the plasma and contribute to the relevant components of the temperature derivative:
\begin{equation}\label{eqn:dTdt}
  \frac{dT_\gamma}{dt} = -3H\,\frac{\displaystyle\rho_\gamma + \rho_e
  + \sum\limits_{x=1}^3\rho_{\nu_x}
  + p_\gamma + p_e +  \sum\limits_{x=1}^3p_{\nu_x}}
  {\displaystyle\frac{d\rho_\gamma}{dT_\gamma} + \frac{d\rho_e}{dT_\gamma}
  + \sum\limits_{x=1}^3\frac{d\rho_{\nu_x}}{dT_\gamma}},
\end{equation}
\noindent where the subscript $e$ denotes the sum of electron and positron components.  In writing Eq. \eqref{eqn:dTdt}, we have omitted the terms associated with baryons for the sake of brevity.  The actual numerical derivative does include baryons and related derivatives.  As each individual neutrino species decouples, we remove it from the summation for $\rho$, $p$, and $d\rho/dT_\gamma$.  Eventually, all three neutrino species decouple and we obtain the default derivative in the WK code. \newline

\noindent The electron neutrino and antineutrino distributions are inputs into the weak interaction rates:
\begin{align}
  n+\nu_e&\leftrightarrow e^- + p,\\
  n+e^+&\leftrightarrow \overline{\nu}_e + p,\\
  n&\leftrightarrow e^- + \overline{\nu}_e + p.
\end{align}

\noindent For this work we alter the $n\rightarrow p$ and $p\rightarrow n$ rates given in the WK code to use the hotter $\nu_e$ and \bnue spectra. The weak interaction rates influence the neutron to proton ratio, $n/p$, which results in the prediction of the primordial $^{4}$He mass fraction:
\begin{equation}\label{He_prod}
  Y_P \equiv \frac{4n_{\rm He}}{n_b} \simeq \frac{4(n_n/2)}{n_p + n_n}=\frac{2n/p}{1+n/p}.
\end{equation}

\noindent Eq. \eqref{He_prod} is only an approximation as not all of the remaining neutrons are incorporated into $^4$He nuclei.  The hotter \nue and \bnue spectra will keep $n/p$ in chemical equilibrium longer, thereby reducing the ratio and $Y_P$.  However, $n/p$ is not solely determined by the \nue and \bnue spectra.  The comparison of the expansion rate $H$ and the $n\leftrightarrow p$ rates determines the epoch of weak freeze--out, and thus $n/p$.  The hotter \nue and \bnue spectra imply a larger neutrino energy density, a faster expansion rate, and an earlier epoch of weak freeze--out.  The earlier epoch leads to a larger $n/p$.  Therefore, lower neutrino decoupling temperatures induce two competing effects with regards to $^4$He production.  We will adopt this theory as our initial paradigm for the behavior of $Y_P$.\newline

\begin{figure}[!ht]
\begin{center}
\includegraphics[scale=0.684]{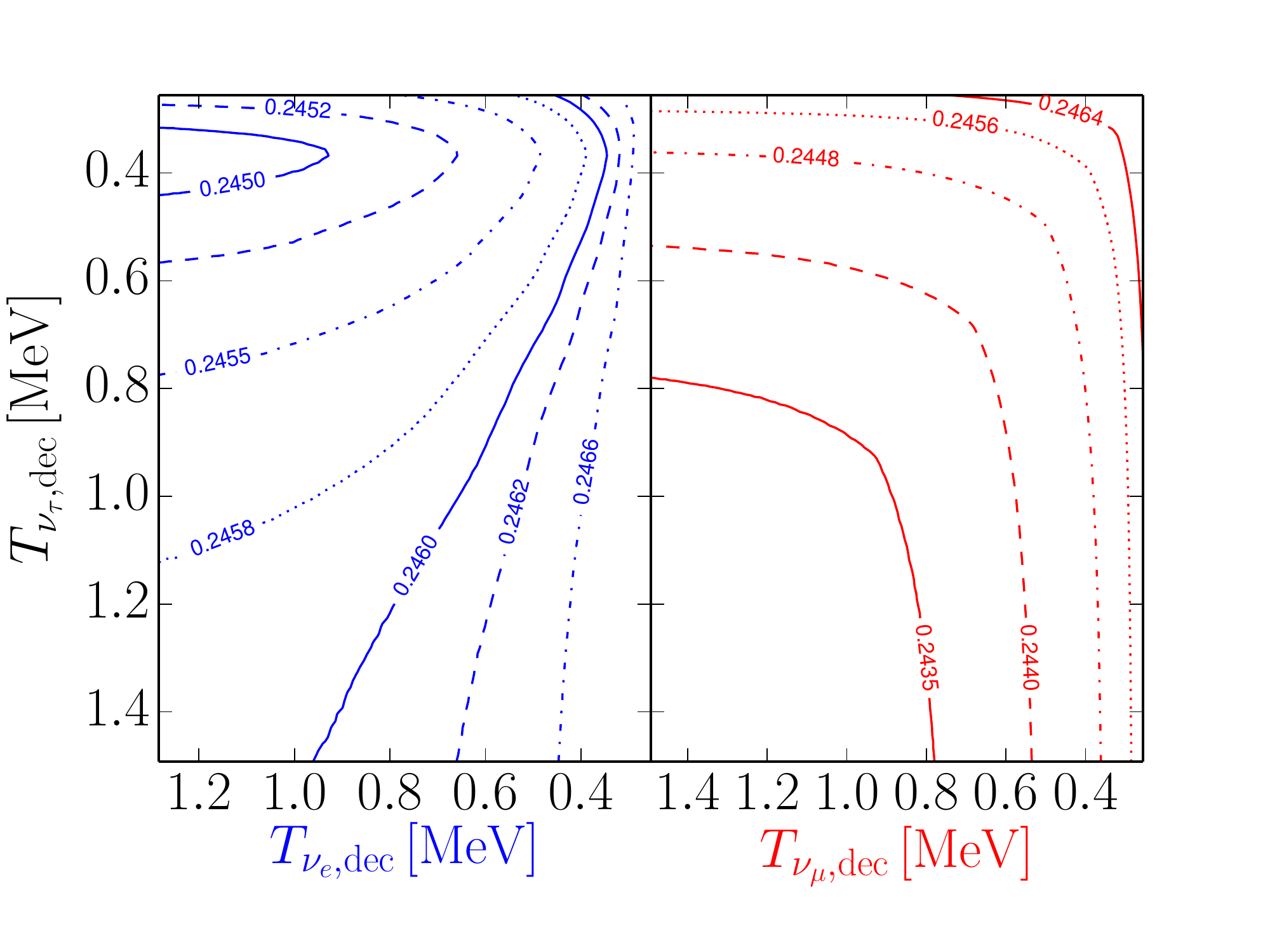}
  \caption{(Color online) {Contours of constant $Y_P$ in the \tdect versus \tdece (left)
  and \tdecm (right) parameter spaces.
  The left panel uses a constant $\tdecm=0.245\,{\rm MeV}$.
  The right panel uses a constant $\tdece=0.245\,{\rm MeV}$.}}
\label{fig:fig5}
\end{center}
\end{figure}

\noindent We demonstrate the interesting behavior of the primordial $^{4}$He mass fraction as a function of neutrino decoupling temperatures in Fig. \ref{fig:fig5}. When holding \tdece fixed, the \tdecm--\tdect parameter space shown in the right panel gives the expected behavior.  A decreasing decoupling temperature for either \num or \nut precipitates an earlier epoch of weak freeze--out, yielding a larger $n/p$ and $Y_P$.  Nucleosynthesis is insensitive to the flavor of neutrinos which are not of \nue type and so the right panel of Fig. \ref{fig:fig5} is symmetric about the line $\tdecm=\tdect$.  No such symmetry exists in the left panel of Fig. \ref{fig:fig5}.  If we hold \tdect fixed and examine the change in $Y_P$, we observe that $Y_P$ increases with decreasing \tdece.  The monotonicity would seem to indicate that the faster expansion rate dominates over the effect of the hotter \nue spectrum on the $n\leftrightarrow p$ rates.  Conversely, if we hold \tdece fixed and examine the change in $Y_P$, we observe that $Y_P$ \emph{decreases} with decreasing \tdect until a turn--around temperature $\tdect\sim0.4\,{\rm MeV}$. The turn--around temperature is independent of \tdece, however, we do notice that we traverse more contours when decreasing \tdect with larger \tdece than we do with smaller \tdece. Once \tdect becomes smaller than the turn--around temperature, $Y_P$ then begins to increase with decreasing \tdect. This behavior implies that the faster expansion rate does not precipitate an earlier epoch of weak freeze--out until \tdect falls below $\sim0.4\,{\rm MeV}$.  This phenomenon is unlike the behavior described earlier for constant \tdect on the left panel, and the behavior as seen on the right panel. We conjecture that our n\"aive paradigm of $n/p$ as a function of only the \nue, \bnue spectra and expansion rate neglects the effect of important nuclear physics rates involving nuclides with mass number two and three.\newline

\noindent We would like to note that constraints from neutrino decoupling are more encompassing than just having implications for neutrino transition moments. The effect of non--standard neutral current (NC) neutrino--electron interactions on the neutrino decoupling temperature in BBN has been previously considered in Ref. \cite{Mangano}. This work found that laboratory constraints on the non--standard NC couplings do not permit such processes to significantly effect cosmological parameters \cite{Mangano}.  Since the neutrino charge radius generates an additive term to the weak neutrino--electron scattering cross section \cite{Vogel}, studies of non--standard NC neutrino couplings also have implications for the magnitude of the effective charge radius in the primordial plasma.  Other electromagnetic properties of neutrinos can be probed through examinations of the magnetic channel. Limits on neutrino magnetic moments from existing experimental neutrino--electron scattering data have been used to obtain bounds on the neutrino electric millicharge \cite{Gninenko} and generic tensoral couplings of neutrinos to charged fermions \cite{Healey} . Since the magnetic interaction coupling is tensoral in nature, it should not be readily assumed that the insensitivity of BBN to non--standard NC couplings translates to the magnetic channel. Non--standard tensoral couplings are being further probed by the TEXONO experiment \cite{DenizNS}, and so a proper treatment of neutrino decoupling could be a complimentary approach to experiment in studying these interactions. \newline

\begin{figure}[!ht]
\begin{center}
\includegraphics[scale=0.72]{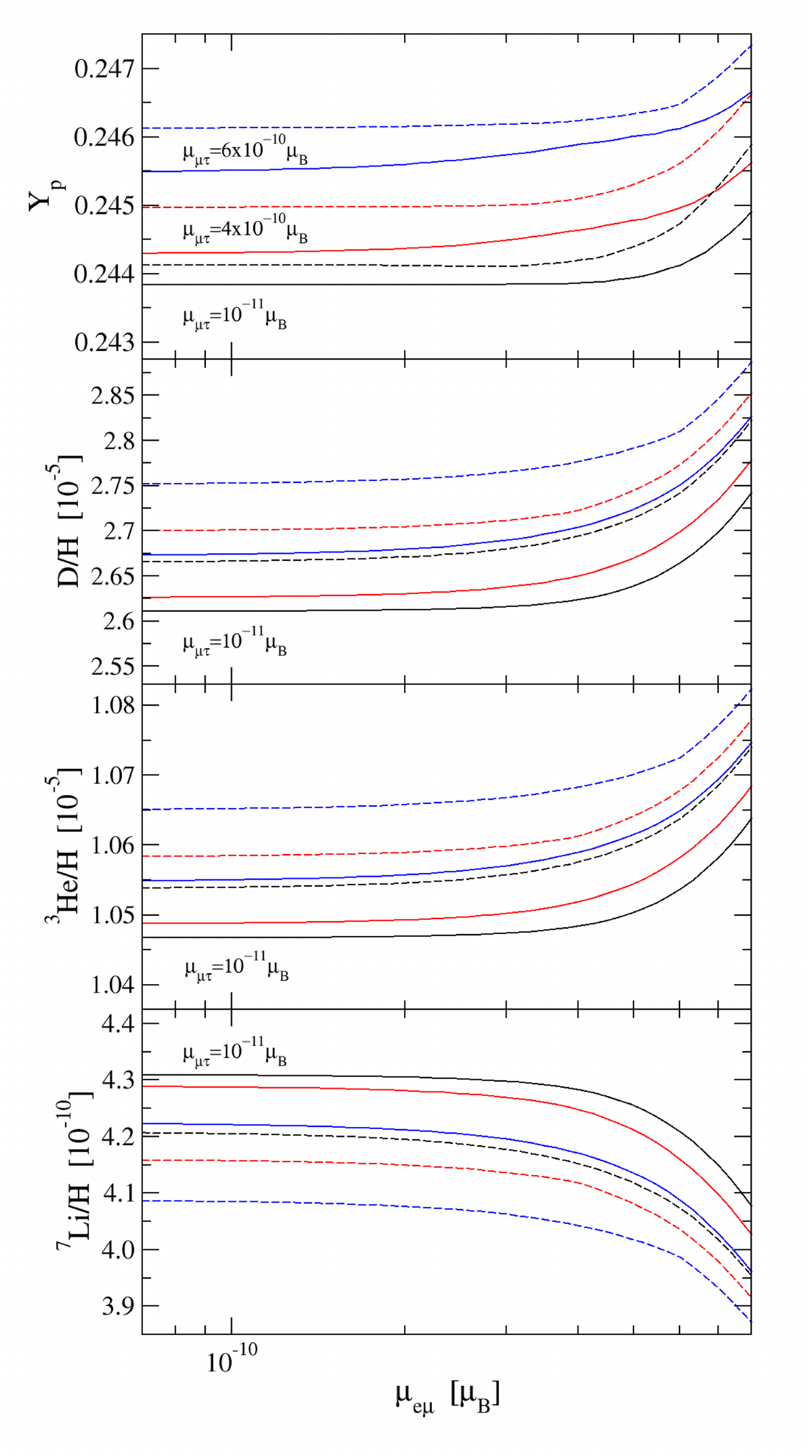}
\caption{(Color online) The change in relative abundances of $^4$He, D, $^3$He, and $^7$Li as a function of the transition neutrino magnetic moment $\mu_{e\mu}$. The solid lines show the results when $\mu_{e\tau}=10^{-11}\mu_B$ (black - $\mu_{\mu\tau}=10^{-11}\mu_B$, red - $\mu_{\mu\tau}=4\times 10^{-10}\mu_B$, and blue - $\mu_{\mu\tau}=6\times 10^{-10}\mu_B$). The dashed lines are the results for $\mu_{e\tau}=6\times 10^{-10}\mu_B$ with the colors representing the same values of $\mu_{\mu\tau}$ that they did in the solid case.}
\label{fig:fig6}
\end{center}
\end{figure}

\noindent Since for this work we are concerned with the connection between neutrino decoupling and transition magnetic moments, we proceed with the assumption that the magnetic interaction is the only new physics affecting the neutrino decoupling temperatures.  In Fig. \ref{fig:fig6} we show the dependence of the primordial abundances on the transition moment $\mu_{e\mu}$.  It is interesting to note that increases in magnetic moment cause a decrease in the predicted lithium abundance but do not have a large enough effect to be a potential solution to the lithium problem since even relatively high magnetic moments $\sim 6.3\times 10^{-10}\mu_B$ can only reduce lithium to $\sim 3.94 \times 10^{-10}$ (far from the observationally inferred value of $(1.6\pm 0.3)\times10^{-10}$ \cite{Ryan}). The known $2\sigma$ tension between the deuterium abundance found with codes such as \texttt{PArthENoPE} and the observationally inferred value \cite{Valentino} is also a factor in this work. With the integrated $n\rightarrow p$ and $p\rightarrow n$ rates, our modified WK code gives a value of ${\rm D/H} = 2.61\times 10^{-5}$ for small neutrino magnetic moment which agrees well with the value given by \texttt{PArthENoPE} of ${\rm D/H} = (2.65\pm0.07)\times 10^{-5}$ (with the updated $d(p,\gamma)^3$He reaction rate) \cite{Valentino}. The values found from quasar absorption--line systems (${\rm D/H}=(2.53\pm 0.04)\times 10^{-5}$) and indirect determinations from CMB data therefore do not allow for this work to place exclusions on Majorana neutrino magnetic moments through observational constraints. The $^4$He abundance is also modified by non--zero neutrino transition moments, however the largest found $^4$He abundance for magnetic moments in the range of $(10^{-11} -  6.3\times 10^{-10})\,\mu_B$ is $Y_P \sim 0.2465$. This is well within the observationally inferred range of $0.2465\pm0.0097$ \cite{AOPS}.  Future observations of $Y_P$ may provide constraints on neutrino transition magnetic moments if errors can be reduced to the 1\% level.  D/H could be probed to the sub-one percent level with thirty--meter class telescopes \cite{TMT, GMT, EELT}.  However, transition magnetic moments would tend to increase D/H, thereby causing more tension with current observations \cite{Cooke} then in the case without magnetic moments.\newline

\begin{figure}[!ht]
\begin{center}
\includegraphics[scale=0.684]{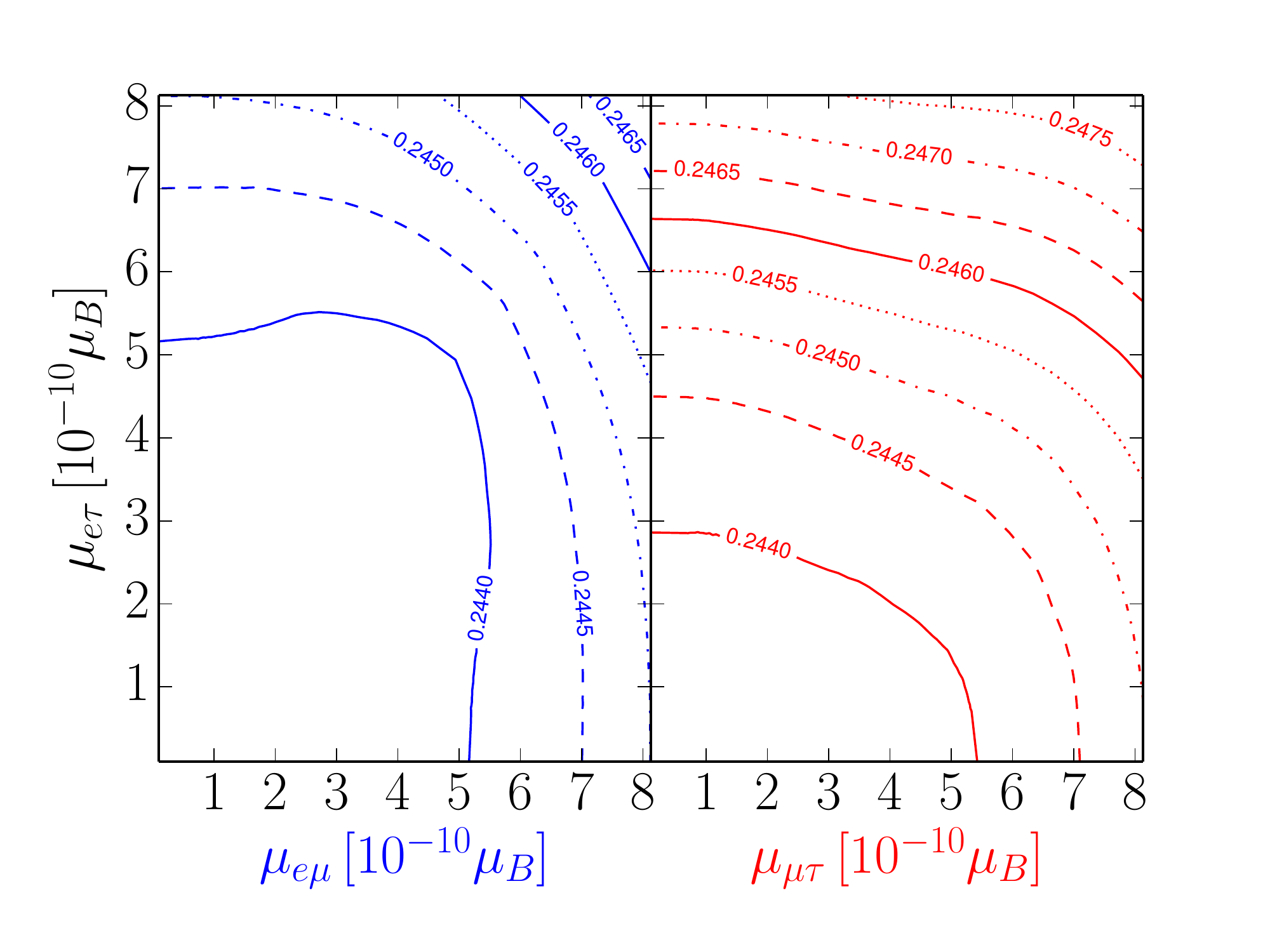}
\caption{(Color online) Contours of constant $Y_P$ in the $\mu_{e\tau}$
  versus $\mu_{e\mu}$ (left) or $\mu_{\mu\tau}$ (right) planes.
  The third transition magnetic moment is set to be $\sim10^{-10}\mu_B$
  in both planes.}
\label{fig:fig7}
\end{center}
\end{figure}

 \noindent In Fig. \ref{fig:fig7} we explore contours of constant $Y_P$ as a function of transition magnetic moments. The effective magnetic moment for electron neutrinos defined in Eq. \eqref{11} is a symmetric function of $\mu_{e\mu}$ and $\mu_{e\tau}$.  Therefore, in the left panel the plot is symmetric about the line $\mu_{e\mu}=\mu_{e\tau}$. The general behavior is that a larger magnetic moment implies a larger value of $n/p$ with a concurrent change in $Y_P$.  However, this is not universally true as the $Y_P=0.2440$ contour does not uphold this trend.  An example of an exception to the general trend occurs for values of $\etau\sim5\times10^{-10}\mu_B$ with $\emu \lesssim 5\times10^{-10}\mu_B$. In this narrow range an increase in \emu results in a decrease in $Y_P$, implying that here the $n\rightarrow p$ rate is faster than $H$. The general trend of larger magnetic moment producing a larger value of $n/p$ seen in the left panel applies to the right panel without exception. For the right panel of Fig. \ref{fig:fig7}, there is no symmetry along the $\etau=\mutau$ line. The asymmetry in the space implies there still exists a competition between the $n\rightarrow p$ rate and $H$, but the effect is not as dramatic as in the parameter space of the left panel.\newline

\begin{figure}[!ht]
\begin{center}
\includegraphics[scale=0.633]{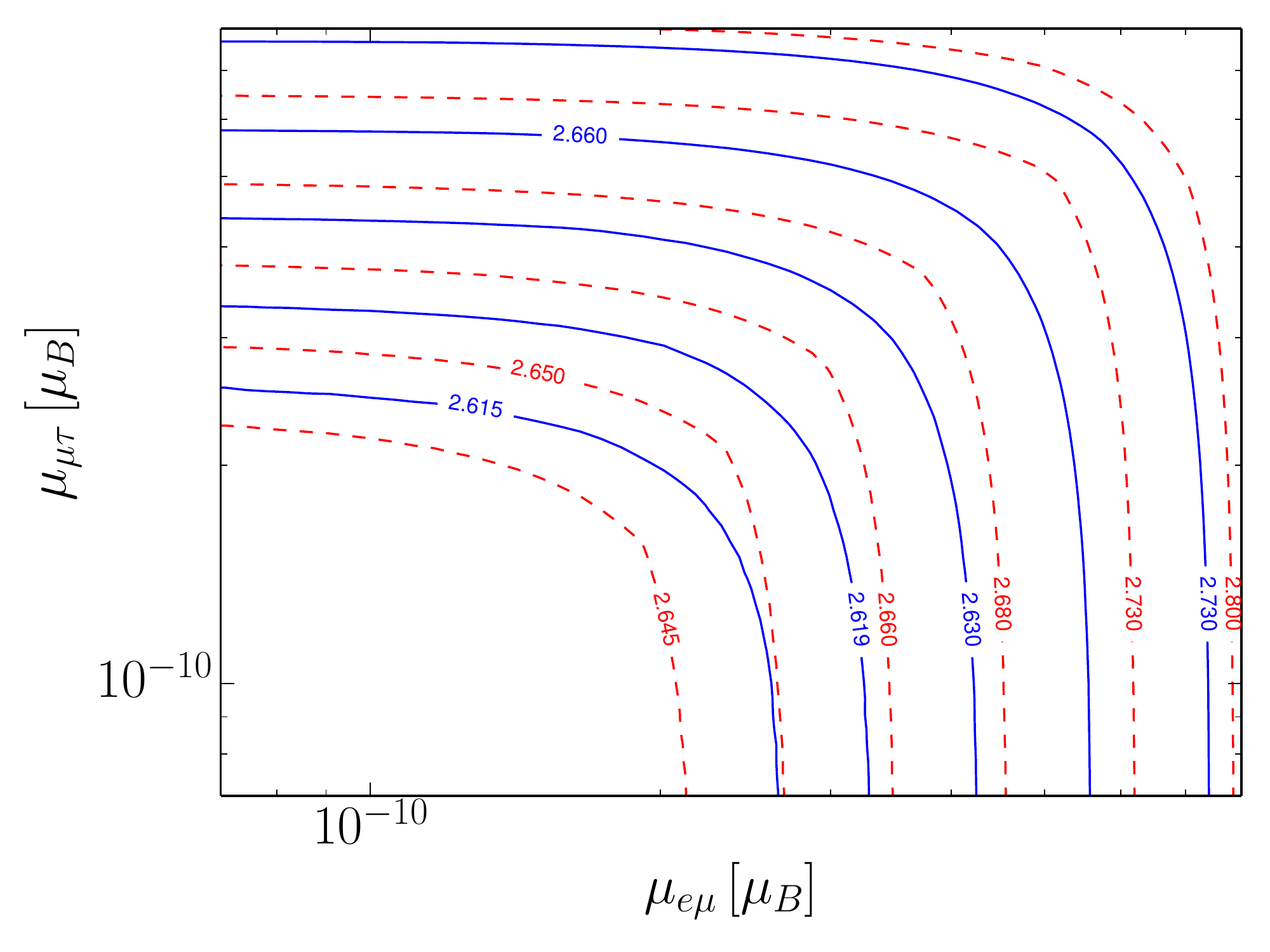}
\caption{(Color online) Contours of constant $10^5\times$ D/H in the \mutau versus
  $\mu_{e\mu}$ plane.  The solid contours correspond to
 $\etau=10^{-10}\mu_B$ and the dashed contours correspond to $\etau=4.9\times10^{-10}\mu_B$.}
\label{fig:fig8}
\end{center}
\end{figure}

\noindent The deuterium abundance is less sensitive than $Y_P$ to the $n/p$ ratio.  Fig. \ref{fig:fig8} shows the primordial relative abundance of deuterium (with respect to hydrogen) multiplied by $10^5$ as a function of neutrino transition moments \emu and \mutau. The different sets of contours correspond to different values of \etau, namely $\etau=10^{-10}\mu_B$ for the solid contours and $\etau= 4.9\times10^{-10}\mu_B$ for the dashed contours.  For both sets of contours, the value of D/H increases more rapidly for increasing \emu as compared to increasing \mutau.  The larger relative change of D/H as compared to $Y_P$ (7\% vs. 1\% over the same parameter space) is not due to the $n/p$ ratio, but instead a result of a larger initial entropy per baryon.  If neutrinos remain coupled to the plasma after the initiation of electron--positron annihilation, then entropy is transferred from the plasma into the neutrino seas. Therefore, a late neutrino--decoupling epoch implies a loss of entropy in the photon sector. Ref.\ \cite{PlanckXVI} gives the baryon density as $\omega_b=0.022068$, which is inversely related to the entropy per baryon. To match the value of $\omega_b$ from Ref.\ \cite{PlanckXVI}, we must begin BBN with a larger entropy per baryon. Fig. \ref{fig:fig8} validates the exquisite sensitivity of the deuterium abundance to the initial entropy per baryon. \newline

\begin{figure}[!ht]
\begin{center}
\includegraphics[scale=0.633]{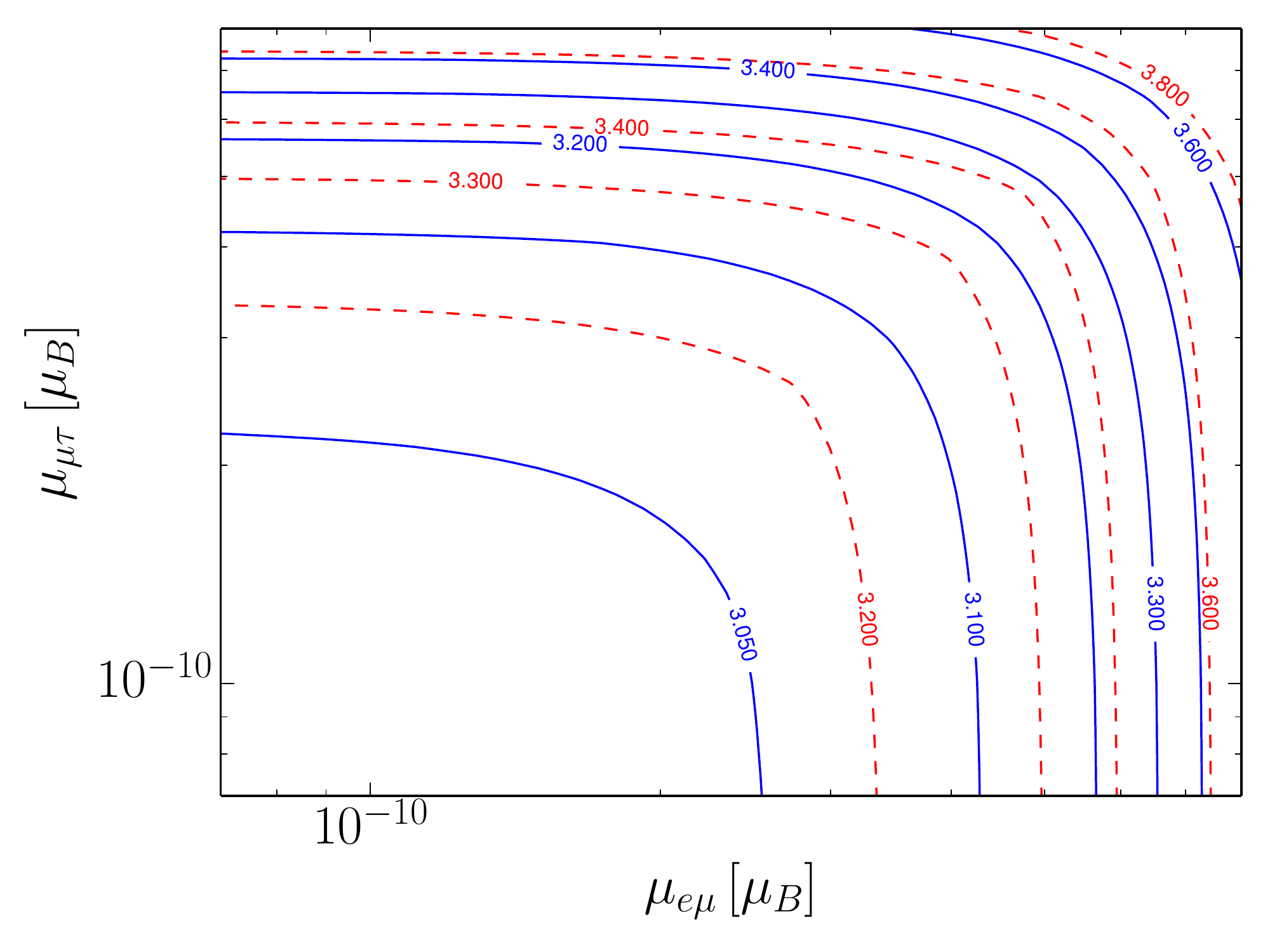}
\caption{(Color online) Contours of constant \neff in the \mutau versus
  $\mu_{e\mu}$ plane.  The solid contours correspond to
  $\etau=10^{-10}\mu_B$ and the dashed contours correspond to
  $\etau=4.9\times10^{-10}\mu_B$.}
\label{fig:fig9}
\end{center}
\end{figure}

\noindent Although examinations of the primordial abundances could not yield approximate constraints on neutrino transition moments, there is an additional parameter with high sensitivity to neutrino decoupling, namely $N_{\text{eff}}$, defined by

\begin{equation}
\label{45}
\rho_{\text{rel}}=\left(1+\frac{7}{8}\left(\frac{4}{11}\right)^{4/3}N_{\text{eff}}\right)\frac{\pi^2}{15}T_{\gamma}^4 ,
\end{equation}

\noindent  where $\rho_{\text{rel}}$ is the radiation energy density, i.e.\ the
sum of the photon and neutrino energy densities. We examine contours of
constant $N_{\text{eff}}$ as a function of neutrino transition moments in
Fig. 9. The axes and contour sets of Fig. \ref{fig:fig9} are the same as
Fig. \ref{fig:fig8}.  As defined in Eq. \eqref{45}, \neff is independent of
neutrino flavor and therefore symmetric and monotonic in any decoupling
temperature parameter space.  However, Fig. \ref{fig:fig9} shows \neff in the
transition magnetic moment space.  The solid blue $\neff=3.050$ contour is not
symmetric in this parameter space.  The asymmetry is the result from different
decoupling temperatures for \nut compared to \nue for equivalent effective
magnetic moments.  The magnetic moment channel conduces a rate similar in value
to the weak scattering rates of \nut as compared to the weak scattering rates
of \nue, as shown in Figs. 2 and 3.  The magnetic channel
contributes more to lowering the decoupling temperature of \nut than it does
for \nue at these particular magnetic moment values.  For larger magnetic
moments, the magnetic channel dominates over the weak channels regardless of
flavor, as seen in Fig. 4.  Therefore, the $\neff>3.100$ contours of
Fig. \ref{fig:fig9} are symmetric.  We note that it is possible to use the
one--sigma range $\neff=3.30\pm0.27$ \cite{PlanckXVI} from the most recent
Planck data to constrain the values of neutrino magnetic moments. We find that
$N_{\text{eff}}$ is able to exclude transition moments larger than $\sim6\times
10^{-10}\mu_B$ to two sigma.  \newline

\section{Conclusions}

Our analysis revealed that Majorana neutrino transition magnetic moments can alter the nucleosynthesis of the primordial elements by keeping the neutrinos coupled during the BBN epoch. Investigating active neutrinos which are Majorana in nature allowed us to explicitly look at the cosmology associated with this physics by quantifying the connections between transition magnetic moment and observables. However, here we were limited by the tendency of BBN codes to oversimplify the neutrino physics. Since a transition magnetic moment necessarily changes flavor and parity, a proper analysis will use the Boltzmann solvers in \texttt{BURST} to account for the out--of--equilibrium neutrino distributions and the spectral swaps associated with changes in flavor. To circumvent these issues and perform a preliminary investigation of neutrino transition moments in BBN, we employed the decoupling temperature approximation to quantify how magnetically enhanced interaction rates would maintain thermal equilibrium between the neutrinos and charged leptons until late epochs. \newline

\noindent We expected the late neutrino decoupling epochs to beget a faster
expansion rate and result in quicker weak freeze--out of BBN events. $^4$He
exhibited the most exotic behavior of any observable quantity we investigated.
The results for $Y_P$ when changing \etau and \mutau invariably showed a larger
$n/p$ and larger $Y_P$ for increasing magnetic moment strength. When
investigating the effect on $Y_P$ with changing \emu and \etau, the larger
magnetic moment did not always engender a larger value of $Y_P$.  However, the
general trend of larger magnetic moments yielding larger values of $n/p$ was
still apparent.  In all cases, the change in $Y_P$ is still consistent with the
observational bounds of Ref. \cite{AOPS}. The trends for D/H, and to a lesser
extent $^3$He/H and $^7$Li/H, follow from changes in the initial entropy.  A
larger initial entropy--per--baryon in the plasma causes increases in deuterium
and $^3$He, and a decrease in $^7$Li. \newline

\noindent Changes in \neff from the resultant
extra energy density of a coupled neutrino limit $\mu_{xy}\lesssim
6\times10^{-10}\,\mu_B$. Our approximate
constraint was limited by current observational errors in $N_{\text{eff}}$
measurements.  Future advances in CMB astronomy could limit the error in \neff
to percent levels, and the baryon density to sub-percent levels. However the projected sensitivities would still only allow for this work to obtain constraints on magnetic moments at the order of $10^{-10}\mu_B$. For instance the projected sensitivity on \neff of $\sigma\sim0.021$ for CMB-S4 \cite{Abazajian201566} applied about the base value found using our modified WK code in the absence of magnetic moments yields an \neff of $3.038\pm 0.021$. This range would imply a transition magnetic moment constraint of $\mu_{xy}\lesssim 2.5\times 10^{-10}\mu_B$ to $2\sigma$. This is still roughly an order of magnitude above the recently found constraints on transition moments using sub--MeV Borexino data \cite{Canas}. If we consider values of magnetic moments closer to the recent laboratory bounds, we find percent changes in D/H to be $0.021\%$ , $Y_P$ to be $0.0021\%$ , and \neff to be $0.068\%$ for the case that all three transition moments are $\sim7\times10^{-11}\mu_B$. BBN observations are not forecast to be at this level of precision in the near future. Thus if experiments such as SPT-3 \cite{2014SPIE.9153E..1PB}, the Simons array
\cite{2014SPIE.9153E..1FA}, or CMB-S4 \cite{Abazajian201566} determine \neff is
statistically larger than the predicted value of 3.046 \cite{neff:3.046}, it is unlikely that beyond--the--standard model neutrino magnetic moments are a contributing factor. Although the calculated abundance changes are below current observational sensitivities, the technique outlined in this paper shows the way toward BBN probes of other beyond--the--standard model neutrino sector issues. \newline

\noindent  Since we find that transition moments begin to play a role in neutrino decoupling at the now excluded order of $10^{-10}\mu_B$, this work suggests that effects on cosmological parameters from neutrino magnetic channels are immaterial. However the instantaneous decoupling approximation used here neglects out--of--equilibrium effects which are known to distort neutrino spectra. The energy dependence of standard weak neutrino interactions implies that the high energy tail of the neutrino distribution interacts most strongly which in turn causes energy dependent spectral distortions. However as demonstrated in section II, the energy dependence of magnetic interactions significantly differs from those of weak interactions (also evident from cross sections in Appendix C). Thus magnetically modified spectral distortions may have more pronounced effects on \neff than what was found here with decoupling considerations alone. Additionally given the sensitivity of $^4$He to electron neutrinos demonstrated by Figs. \ref{fig:fig5} and \ref{fig:fig7}, changes in the electron neutrino number density from spectral swaps associated with flavor changing currents should be carefully considered. In order to fully understand possible effects of neutrino magnetic moment in BBN, the inclusion of neutrino spectral distortions and swaps deserves to be studied by a more rigorous treatment.  \newline

\acknowledgments

NV would like to thank the organizers of the TALENT NT4A summer school Richard Cyburt, Morten Hjorth-Jensen, and Hendrik Schatz for encouraging the pursuit of this work. We thank Joe Kapusta, Charles Gale, Yamac Pehlivan, Chad Kishimoto, Mark Paris and Pat Diamond for useful discussions. We also thank the referee for useful comments. We would like to acknowledge the Institutional Computing Program at Los Alamos National Laboratory for use of their HPC cluster resources. This work was supported in part by U.S. National Science Foundation Grants No. PHY-1205024 and No. PHY-1514695 at the University of Wisconsin, No. PHY-1307372 at UC San Diego, and in part by the University of Wisconsin Research Committee with funds granted by the Wisconsin Alumni Research Foundation.

\section{Appendix A: Inverse Screening Length}

Following the discussion given by Kapusta and Gale \cite{Kapusta} the exact inverse screening length is found from the longitudinal polarization in the static infrared limit and is given in natural units by 

\begin{equation}
\label{15}
k_{sc}^2 = e^2 \frac{\partial n}{\partial \mu},
\end{equation}

\noindent where we must use the net electron density $n=n_e\equiv n_{e^-}-n_{e^+}$ given by

\begin{equation}
\label{16}
n_e = \frac{1}{\pi^2}\int_{0}^{\infty} dp\, p^2 \left[\frac{1}{e^{(E-\mu)/T}+1}-\frac{1}{e^{(E+\mu)/T}+1}\right]
\end{equation} 

\noindent because we have an electron--positron plasma \cite{Braaten}. For the case of zero chemical potential

\begin{equation}
\label{17}
\frac{\partial n_e}{\partial \mu}\, \bigg|_{\mu\rightarrow 0} = \frac{2}{\pi^2}\frac{1}{T}\int_{0}^{\infty} dp\,p^2 \frac{e^{E/T}}{\left(e^{E/T}+1\right)^2}.
\end{equation}

\noindent Then the inverse screening length is

\begin{equation}
\label{18}
k_{sc}^2 = \frac{4\alpha}{\pi T}\int_0^{\infty} dp\,p^2 \frac{1}{1+\cosh(E/T)}.
\end{equation}

\noindent Note that in the relativistic limit $T>>m_e$ the above inverse screening length reproduces the well known relation $\omega_{p}^2 = \frac{1}{3}k_{sc}^2$ \cite{Kapusta}, where the plasma frequency in the relativistic limit is given by $\omega_p^2 = \frac{4\alpha\pi}{9} T^2$ at zero chemical potential \cite{Braaten}.

\section{Appendix B: Interaction Rates with Fermi--Dirac Distributions}

 Gondolo and Gelmini \cite{Gondolo} outline the procedure for turning the integral of Eq. \eqref{4} over $d^3 p_1$ and $d^3 p_2$ into an integral over the Mandelstam variable $s$. Using the change of variables 

\begin{eqnarray}
\label{5}
E_{+}&=&E_{1}+E_{2}, \\ \nonumber 
E_{-}&=&E_{1}-E_{2}, \\ \nonumber
\end{eqnarray}

\noindent gives $d^3 p_1 d^3 p_2 = 2\pi^2 E_1 E_2 dE_+ dE_- ds$  and can write

\begin{equation}
\label{6}
\frac{1}{1+e^{E_1 /T}}\frac{1}{1+ e^{E_2 /T}} = \frac{e^{-\frac{E_+}{2T}}}{2}\frac{1}{\cosh\left(\frac{E_+}{2T}\right)+\cosh\left(\frac{E_ -}{2T}\right)}.
\end{equation}

 \noindent For massless neutrinos interacting via the annihilation channel the numerator of Eq. \eqref{4} becomes

\begin{eqnarray}
\label{7}
\int \sigma v_{\text{Mol}} \frac{d^3 p_1}{1+e^{E_1 /T}}\frac{d^3 p_2}{1+e^{E_2 /T}}  = \frac{\pi^2}{2}  \int_{4m_e^2}^{\infty} \sigma\, s \, ds \int_{\sqrt{s}}^{\infty} dE_{+} e^{-\frac{E_{+}}{2T}}\int_{-\sqrt{E_+^2 -s}}^{\sqrt{E_+^2 -s}}  \frac{dE_-}{\cosh\left(\frac{E_+}{2T}\right)+\cosh\left(\frac{E_-}{2T}\right)}.
\end{eqnarray}

\noindent The last integral of Eq. \eqref{7} is known analytically to be \cite{Gradshteyn}

\begin{equation}
\label{8}
\int \frac{dx}{\cosh(a)+\cosh(x)} = \text{cosech}(a) \left[\text{ln}\cosh\left(\frac{x+a}{2}\right)-\text{ln}\cosh\left(\frac{x-a}{2}\right)\right],
\end{equation}

\noindent which allows the thermally averaged cross section for the annihilation channel to be reduced to a two dimensional integral and written as

\begin{equation}
\label{appc:9}
\left<\sigma v_{\text{Mol}}\right>_{\text{ann}} = \frac{4\pi^2 T^2}{n_\nu^2}\int_{4m_e^2} ^\infty \sigma s ds \int_{\sqrt{s}/T} ^\infty dx \frac{e^{-x}}{1-e^{-x}}\left[\frac{\sqrt{x^2 -s/T^2}}{2}+\ln\left(\frac{1+e^{-\frac{x+\sqrt{x^2 - s/T^2}}{2}}}{1+e^{-\frac{x-\sqrt{x^2 - s/T^2}}{2}}}\right)\right],
\end{equation}

\noindent which is the expression given in Eq. \eqref{9}. For the scattering channel we obtain

\begin{align}
\label{appc:10}
  \left<\sigma v_{\text{Mol}}\right>_{\text{scatt}} = \frac{4\pi^2 T^2}{n_\nu n_e}
  &\int_{2m_e^2} ^\infty \sigma \left(s-m_e^2 \right) ds
  \int_{\sqrt{s}/T} ^\infty dx \frac{e^{-x}}{1-e^{-x}}\\
  \nonumber\times &\left[\frac{\sqrt{x^2 -s/T^2}\sqrt{1-2m_e^2 /s}}{2}
  +\ln\left(\frac{1+e^{-\frac{x+\sqrt{x^2 - s/T^2}\sqrt{1-2m_e^2 /s}}{2}}}
  {1+e^{-\frac{x-\sqrt{x^2 - s/T^2}\sqrt{1-2m_e^2 /s}}{2}}}\right)\right] ,
\end{align}

\noindent which is the expression given in Eq. \eqref{10}. The kinematic conditions needed to derive the interaction rate for the annihilation channel given by Eq. \eqref{9} are $\left|E_{-}\right|\le \sqrt{E_{+}^2 -s}$ and $E_1 E_2 v_{\text{Mol}} = \frac{1}{2}s$ for massless neutrinos. The results for the scattering channel given by Eq. \eqref{10} make use of the kinematic scattering conditions $\left|E_{-}\right|\le \sqrt{E_{+}^2 -s}\sqrt{1-\frac{2m_e^2}{s}}$ and $E_1 E_2 v_{\text{Mol}} = \frac{1}{2}\left(s-m_e^2\right)$. Note that in using Eq. \eqref{9} and \eqref{10} for rate calculations the phase space integrals $d^3 p_3$ and $d^3 p_4$ contained in the cross section must be multiplied by $g_3$ and $g_4$ (for the internal degrees of freedom of the outgoing particles). The results for all rates in this work have included these multiplicative factors. \newline

\section{Appendix C: Cross Sections}

\noindent The following cross sections assume the incoming neutrino to be polarized. Polarization makes use of the projection operator $\frac{1}{2} \left(1-\gamma_5\right)$ and ultimately leads to a factor of $1/2$ in the magnetic cross sections. However it should be noted that this polarization does not modify the results for the weak interaction due to the structure of the weak interaction vertex.

\subsection{Neutrino--electron/positron scattering cross sections: $\nu\left(p_{1}\right) + e^{\pm}\left(p_{2}\right) \rightarrow \nu\left(p_{4}\right) + e^{\pm}\left(p_{3}\right)$ }
\subsubsection{Weak $\nu - e^-$ Scattering}

\noindent The invariant amplitude squared is

\begin{eqnarray}
\label{19}
\langle|\mathcal{M}_{\text{weak}}|^2 \rangle &=& 16 G_F^2 \Big[\left(g_{A}^2 - g_{V}^2 \right)m_e^2 \left(p_1 \cdot p_4 \right) + \left(g_{A}+g_{V}\right)^2 \left(p_1 \cdot p_2 \right)\left(p_3 \cdot p_4 \right) \\ \nonumber &&\,\,\,\,\,\,\,\,\,\,\,\,\,\,\,\,\,+ \left(g_A -g_V \right)^2 \left(p_2 \cdot p_4 \right)\left(p_1 \cdot p_3 \right) \Big] \\ \nonumber & = & 4 G_F^2 \left[2\left(g_{V}^2 - g_{A}^2 \right)m_e^2 t + \left(g_{A}+g_{V}\right)^2 \left(s-m_e^2 \right)^2 + \left(g_A -g_V \right)^2 \left(s+t-m_e^2 \right)^2 \right],
\end{eqnarray}

\noindent where for electron neutrinos $g_A = \frac{1}{2}\, ,\, g_V = \frac{1}{2} + 2\sin^2\theta_w$ and for muon/tau neutrinos $g_A = -\frac{1}{2}\, ,\, g_V = -\frac{1}{2} + 2\sin^2\theta_w$. The Mandelstam variables in the second expression of Eq. \eqref{19} are defined by

\begin{eqnarray}
\label{20}
t &=& q^2 = \left(p_{1}-p_{4}\right)^2 = -2p_{1}\cdot p_{4} = 2p_1 \cdot p_3 - 2p_1 \cdot p_2 ,\\ \nonumber
s &=& \left(p_{1}+p_{2}\right)^2 = m_{e}^2 + 2p_{1}\cdot p_{2} .
\end{eqnarray}

 \noindent Then the cross section can be found to be

\begin{equation}
\label{21}
\sigma(s) =  \frac{G_F^2}{4\pi}\frac{\left(s-m_e^2\right)^2}{s}\left[\left(g_{A}+g_{V}\right)^2 + 2g_A \left(g_A -g_V \right)\frac{m_e^2}{s} + \frac{1}{3}\left(g_A -g_V \right)^2 \left(1-\frac{m_e^2}{s}\right)^2 \right]
\end{equation}

\noindent after the differential cross section has been integrated from $t_{\text{min}} = -\frac{\left(s-m_e^2 \right)^2}{s}$ to $t_{\text{max}} = 0$. 

\subsubsection{Weak $\nu - e^+$ Scattering}

\noindent Making use of crossing symmetry with $p_{2}\leftrightarrow p_{3}$ have

\begin{eqnarray}
\label{22}
\langle|\mathcal{M}_{\text{weak}}|^2 \rangle &=& 16 G_F^2 \Big[\left(g_{A}^2 - g_{V}^2 \right)m_e^2 \left(p_1 \cdot p_4 \right) + \left(g_{A}+g_{V}\right)^2 \left(p_1 \cdot p_3 \right)\left(p_2 \cdot p_4 \right) \\ \nonumber &&\,\,\,\,\,\,\,\,\,\,\,\,\,\,\,\,\,+ \left(g_A -g_V \right)^2 \left(p_3 \cdot p_4 \right)\left(p_1 \cdot p_2 \right) \Big] \\ \nonumber & = & 4 G_F^2 \left[2\left(g_{V}^2 - g_{A}^2 \right)m_e^2 t + \left(g_{A}+g_{V}\right)^2 \left(s+t-m_e^2 \right)^2 + \left(g_A -g_V \right)^2 \left(s-m_e^2 \right)^2 \right] ,
\end{eqnarray}

\noindent and so the cross section is

\begin{equation}
\label{23}
\sigma(s) =  \frac{G_F^2}{4\pi}\frac{\left(s-m_e^2\right)^2}{s}\left[\left(g_{A}-g_{V}\right)^2 + 2g_A \left(g_A +g_V \right)\frac{m_e^2}{s} + \frac{1}{3}\left(g_A +g_V \right)^2 \left(1-\frac{m_e^2}{s}\right)^2 \right].
\end{equation}

\subsubsection{Magnetic}

 \noindent It is well known that the differential cross section for Coulomb scattering diverges at $t=q^2 \rightarrow 0$. Here we will implement a cutoff which can be generically written as $t_{\text{max}}\rightarrow -|t_{\text{max}}|$. The averaged matrix element squared is given by

\begin{eqnarray}
\label{24}
\langle|\mathcal{M_\gamma}|^2 \rangle & = & \frac{32 \pi^2 \alpha^2 \mu_{\nu}^{2}}{m_{e}^2}\frac{1}{p_1 \cdot p_4}\left(p_1 \cdot p_2 \right)\left( p_1 \cdot p_3\right) \\ \nonumber
& = & \frac{4 e^2 \kappa_{\nu}^2}{t}\left(m_{e}^2 -s\right)\left(s+t-m_{e}^2\right) ,
\end{eqnarray}

\noindent where $\kappa_\nu=\mu_\nu \,\mu_B$. The cross section is

\begin{equation}
\label{25}
\sigma(s) =  \frac{\pi \alpha^2 \mu^2_{\nu}}{m_e^2}\left[\frac{|t_{\text{max}}|}{s-m_e^2}-\frac{s-m_e^2 }{s}+\ln{\frac{\left(s-m_e^2\right)^2}{s\, |t_{\text{max}}|}}\right] .
\end{equation} 

 \noindent The cross section for neutrino--electron scattering applies to neutrino--positron scattering as well since using crossing symmetry with $p_{2}\leftrightarrow p_{3}$ can see from Eq. \eqref{24} that the invariant amplitude squared is symmetric under this interchange. \newline

\subsection{$\nu$$\bar{\nu}$ annihilation cross sections: $\nu\left(p_{1}\right) + \bar{\nu}\left(p_{2}\right) \rightarrow e^{-}\left(p_{3}\right) + e^{+}\left(p_{4}\right)$ }

\subsubsection{Weak}

\noindent For this process the invariant amplitude squared is

\begin{eqnarray}
\label{26}
\langle|\mathcal{M}_{\text{weak}}|^2 \rangle &=& 16 G_F^2 \Big[\left(g_{V}^2 - g_{A}^2 \right)m_e^2 \left(p_1 \cdot p_2 \right) + \left(g_{A}+g_{V}\right)^2 \left(p_1 \cdot p_4 \right)\left(p_3 \cdot p_2 \right) \\ \nonumber && \,\,\,\,\,\,\,\,\,\,\,\,\,\,\,\,\,+ \left(g_A -g_V \right)^2 \left(p_2 \cdot p_4 \right)\left(p_1 \cdot p_3 \right) \Big] \\ \nonumber & = & 4 G_F^2 \left[2\left(g_{V}^2 - g_{A}^2 \right)m_e^2 s + \left(g_{A}+g_{V}\right)^2 \left(t-m_e^2 \right)^2 + \left(g_A -g_V \right)^2 \left(s+t-m_e^2 \right)^2 \right] ,
\end{eqnarray}

\noindent where the first expression of Eq. \eqref{26} can be obtained from a $p_2 \leftrightarrow p_4\,, m_e^2 \rightarrow -m_e^2$ switch to the scattering invariant amplitude given by the first expression of Eq. \eqref{19}. Use of this switch requires that the annihilation process is then defined by $\nu\left(p_{1}\right) + \bar{\nu}\left(p_{2}\right) \rightarrow e^{-}\left(p_{3}\right) + e^{+}\left(p_{4}\right)$ with four--momentum conservation $p_1 + p_2 = p_3 + p_4$. The second expression of Eq. \eqref{26} comes from a $s\leftrightarrow t$ switch in the second expression of Eq. \eqref{19} where here
\begin{eqnarray}
\label{27}
s &=& q^2 = \left(p_{1}+p_{2}\right)^2 = 2p_{1}\cdot p_{2} = 2p_1 \cdot p_3 + 2p_1 \cdot p_4 , \\ \nonumber
t &=& \left(p_{1}-p_{4}\right)^2 = m_{e}^2 - 2p_{1}\cdot p_{4} ,
\end{eqnarray}

\noindent and so the cross section can be found to be

\begin{equation}
\label{28}
\sigma(s) =  \frac{G_F^2}{2\pi}\sqrt{1-\frac{4m_e^2}{s}}\left[\left(g_V^2 -g_A^2 \right)m_e^2 +\frac{1}{3}\left(g_A^2 + g_V^2 \right)\left(s-m_e^2\right) \right] ,
\end{equation}

\noindent after the differential cross section has been integrated from $t_{\text{min}}=\frac{1}{2}\left[2m_e^2 -s -\sqrt{s\left(s-4m_e^2\right)}\,\right]$ to $t_{\text{max}}=\frac{1}{2}\left[2m_e^2 -s +\sqrt{s\left(s-4m_e^2\right)}\,\right]$.

\subsubsection{Magnetic}

\noindent The invariant amplitude squared is 
\begin{eqnarray}
\label{29}
\langle|\mathcal{M_\gamma}|^2 \rangle & = & \frac{64 \pi^2 \alpha^2 \mu_{\nu}^{2}}{m_{e}^2}\frac{1}{p_1 \cdot p_2}\left(p_1 \cdot p_4 \right)\left( p_1 \cdot p_3\right) \\ \nonumber
& = & \frac{8 e^2 \kappa_{\nu}^2}{s}\left(m_{e}^2 -t\right)\left(s+t-m_{e}^2\right) ,
\end{eqnarray}

\noindent where the above includes the factor of 1/2 due to polarization of the incoming neutrino and antineutrino. And so the cross section is

\begin{equation}
\label{30}
\sigma(s) =  \frac{2 \pi \alpha^2 \mu^2_{\nu}}{6 m_e^2}\sqrt{1-\frac{4m_e^2}{s}}\left(1+\frac{2m_e^2}{s}\right) .
\end{equation}


\end{document}